\documentclass[aps,twocolumn,showpacs,preprintnumbers,amsmath,amssymb]{revtex4}
\usepackage{graphicx}
\begin{document}
%
%
\title{Temperature dependent molecular dynamic simulation of friction}
\author{R.A. Dias}\email{radias@fisica.ufmg.br}
\affiliation{Laborat\'orio de Simula\c{c}\~ao - Departamento de
F\'{\i}sica - ICEX - UFMG 30123-970 Belo Horizonte - MG, Brazil}
\author{M. Rapini} \email{mrapini@fisica.ufmg.br}
\affiliation{Laborat\'orio de Simula\c{c}\~ao - Departamento de
F\'{\i}sica - ICEX - UFMG 30123-970 Belo Horizonte - MG, Brazil}
%
%
%
\author{P.Z. Coura} \email{pablo@fisica.ufjf.br}
\affiliation{Laborat\'orio de Simula\c{c}\~ao - Departamento de
F\'{\i}sica - ICEX - UFMG 30123-970 Belo Horizonte - MG, Brazil}
\author{B.V. Costa} \email{bvc@fisica.ufmg.br}
\affiliation{Laborat\'orio de Simula\c{c}\~ao - Departamento de
F\'{\i}sica - ICEX - UFMG 30123-970 Belo Horizonte - MG, Brazil}
\begin{abstract}
In this work we present a molecular dynamics simulation of a $FFM$
experiment. The tip-sample interaction is studied by varying the
normal force in the tip and the temperature of the surface. The
friction force, $cA$, at zero load and the friction coefficient,
$\mu$, were obtained. Our results strongly support the idea that
the effective contact area, $A$, decreases with increasing
temperature and the friction coefficient presents a clear
signature of the premelting process of the surface.
\end{abstract}

\maketitle
%
%
\section{Introduction}
Friction is one of the oldest phenomenon studied in natural
sciences. In a macroscopic scale it is known that the friction
force between surfaces satisfies the following rules: (1) The
friction is independent of contact area between surfaces; (2) It
is proportional to the normal force applied between surfaces and
(3) The kinetic friction force is independent of relative speed
between surfaces\cite{Nanoscience}. Considering that friction is
the result of many microscopic interactions between the building
atoms at the surfaces, it must depend on factors as roughness ,
temperature and the energy dissipation mechanism at the surfaces.
Therefore, to understand its macroscopic behavior it is necessary
to understand in details the dynamics of interaction between atoms
in the surfaces in contact. In 1987, C. M. Mat et al\cite{Mate}
have used, for the first time, the Friction Force Microscope (FFM)
to investigate friction in nano-scale. That kind of microscope
allows the experimentalist to produce essentially a single contact
between a sharp tip, of atomic dimensions, and the surface.
\cite{Fujisawa,Thomas}.
\begin{figure}[htbp]
\includegraphics[width=9.0cm]{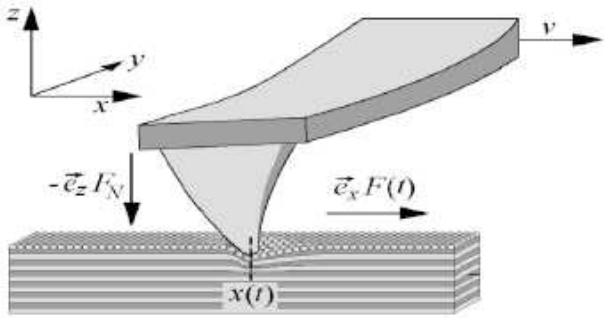}\\
\caption{\scriptsize{Schematic view of a FFM apparatus.
\cite{PReimann}}}\label{FFMesqArt}
\end{figure}
Its basic principle is shown in figure \ref{FFMesqArt}. In this
apparatus the tip stays in touch with the surface while it moves
at constant velocity, $v$, parallel to the surface. The resulting
force in the cantilever makes a torsion on it. This torsion can be
observed by optical techniques. One of the more striking effects
observed is the stick-slip phenomenon: The friction force plotted
as a function of time exhibits a sawtooth behavior.
\cite{Gyalog,Volmer}.

\begin{figure}[htbp]
\includegraphics[width=9.0cm]{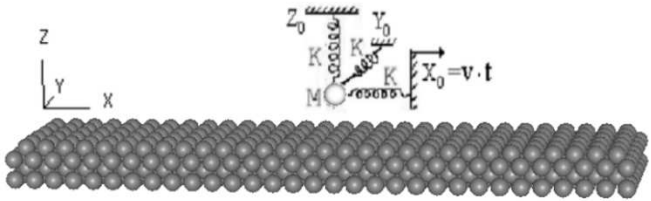}\\
\caption{\scriptsize{Schematic view of our Friction Force apparatus.
}}\label{FFMesqSim}
\end{figure}

Analytically, the motion of the tip can be modelled as follows.
Forces in the tip are a sum of two terms: An interaction force
between the tip and the surface due to the periodic atomic
arrangement of the lattice and a force due to the cantilever. Some
authors by using this approach were able to reproduce several
features of the friction at a nanoscopic scale
\cite{Gnecco,Reimann}. In this work we use molecular dynamics (MD)
simulation to study the friction phenomenon at the atomic scale.
In figure \ref{FFMesqSim} we show a schematic view of the model we
have used in our simulation to reproduce the $FFM$ mechanism
(Figure \ref{FFMesqArt}). The tip is represented by a single
particle that interacts with the surface through a convenient
potential. The springs represent the mechanism we have used to
vary the normal force ($z$ direction) and to measure the lateral
force ($x$ and $y$ directions). By measuring both forces it is
possible to study the friction force behavior under several
circumstances. \\

In a recent work Resende and Costa \cite{flavio1} using molecular
dynamic simulation have studied the migration of an individual atom
on the surface of a $12-6$ Lennard-Jones bcc crystal. They argued
that an observed anomaly occurring in the diffusion constant can be
the signature of a pre-melting process. The migration of an ad-atom
at the surface may occurs by three mechanisms. At low temperature
the adsorbed particle can move through channels on the surface since
thermal motion of atoms at the surface have low amplitude. Once
temperature rises we reach an intermediate state. The surface starts
to melt so that the channels are closed and ad-atoms are stuck in
the vicinity of a surface atom. The situation persists until the
ad-atom is thermally activated and random-walk diffusion occurs. In
summary, the diffusion constant should present a minimum at the
intermediate region. Under the point of view of friction we may ask
what is the effect of this phenomenon over friction. For two
macroscopic sliding surfaces we may not expect to distinguish the
first two process since the contact area is large compared to
interatomic distance. As temperature rises the surface is lubricated
by melted atoms, we may expect a smaller friction coefficient. The
situation is quite different for a small tip in contact with the
surface.

In the following we describe a series of molecular dynamics computer
simulation of the interaction of a small tip with a surface.
Computer simulations give us a very convenient instrument to study
tribological processes. They allow controlled ``numerical
experiments'' where the geometry, sliding conditions and
interactions between atoms can be varied at will to explore their
effect on friction, lubrication, wear and to reach unaccessible
regions for experimentalists. Unlike laboratory experiments,
computer simulations enable us to follow and analyze the full
dynamics of all atoms.  A wide range of potentials have been
employed in numerical studies of tribology. For example, simulations
of metals frequently use the embedded atom method, while studies of
hydrocarbons use potentials that include bond-stretching and so on.
We will not concentrate ourselves in any specific material rather in
general aspects of the problem under consideration. Due to the
general character of our study we content ourselves by using the
Lennard-Jones ($6-12$) potential which is commonly used in studies
of general behavior. We will consider that the interaction of the
tip with the surface does not disturb very much the electronic
structure of the system. This consideration means that we do not
have to deal with the quantum aspects of the problem. This
simplification let us save a lot of computer time. If, for one side,
we lose details of the considered phenomenon, we gain in that we can
study true dynamical and temperature dependent models.

This work is organized as follows. In the section \ref{section2} we
introduce general aspects of the numerical method used, in the
section \ref{section3} we present our results and in section
\ref{section4} we discuss and present some conclusions.
%
%
\section{Simulation background \label{section2}}
Our simulation is carried out by using molecular dynamics (MD)
simulation. A schematic view of the simulation arrangement we have
used is shown in figure \ref{FFMesqSim}. Three springs of elastic
constants $k_x$, $k_y$ and $k_z$ are attached to the particle $M$
that represents a tip point. This arrangement allow us to measure
normal ($F_z$) and parallel ($F_x$, $F_y$) forces on $M$. The
surface is represented by an arrangement of particles which interact
with each other and with the mass $M$ through a truncated
Lennard-Jones ($6-12$) (LJ) potential
\begin{widetext}
\begin{equation}
\Phi_{i,j}(r_{i,j}) = \left \{
\begin{array}{lcc}
  \phi_{i,j}(r_{i,j}) - \phi_{i,j}(r_{c}) - (r_{i,j} - r_{c}) \left(
  \frac{\partial\phi_{i,j}(r_{i,j})}{\partial r_{i,j}}
  \right)_{r_{i,j}=r_c} & ~~~ if & r_{i,j} < r_{c} \\
  0 & ~~~ if & r_{i,j} > r_{c} \\
\end{array} \right.
\label{equation1}
\end{equation}
\end{widetext}
where $\phi_{i,j}(r_{i,j})$ is the complete LJ potential,
\begin{equation}
\phi_{i,j}(r_{i,j}) =4\epsilon_{i,j} \left[ \left(
\frac{\sigma_{i,j}}{r_{i,j}} \right)^{12} - \left(
\frac{\sigma_{i,j}}{r_{i,j}} \right)^{6} \right].
\end{equation}
The indexes $i$ and $j$ stands for position vectors
$\overrightarrow{r_{i}}$ and $\overrightarrow{r_{j}}$ respectively,
and $1 \leq i \leq N$, where $N$ is the total number of particles
and $r_{i,j} = \left| r_{j}-r_{i} \right|$. A cutoff, $r_{c}$, is
introduced in the potential in order to accelerate the simulation.
If the force on a particle is found by summing contributions from
all particles acting upon it, then this truncation limits the
computation time to an amount proportional to the total number of
particles $N$. Of course, this truncation introduces discontinuities
both in the potential and the force. To smooth these discontinuities
we introduce the constant term $\phi(r_c)$. Another term $\left(
{\partial\phi_{i,j}(r_{i,j})}/{\partial r_{i,j}}
\right)_{r_{i,j}=r_c}$ is introduced to remove the force
discontinuity. Particles in the simulation move according to
Newton's law of motion, which generates a set of $3N$ coupled
equations of motion  which are solved by increasing forward in time
the physical state of the system in small time steps of size $\delta
t$. The resulting equations are solved by using Beeman's method of
integration\cite{Allen,Rapaport,Beeman,Berendsen}. The system is
arranged in 4 layers with free boundary conditions in all
directions. The first layer is frozen in a regular arrangement as in
the $(001)$ surface of a Lennard-Jones bcc crystal in order to
maintain the whole structure as flat as possible.

With the tip far away from the surface we thermalize the system at
temperature $T$. After thermalization, the tip is pushed in a
direction parallel to the surface at constant velocity $v_p$. For
each simulation the distance between the spring and the surface is
fixed at the start, so that we can control the perpendicular force
on the tip. By measuring the size variation of the springs we can
calculate the lateral, $F_x$, and the perpendicular force, $F_z$, on
the tip.
The temperature, $T$, of the surface can be controlled by using a
velocity renormalization scheme (See for example \cite{pablo1} and
references therein). From the equipartition theorem we can write
that
\begin{equation}
\langle v^2 \rangle = 3\frac{k_B}{m}T.
\label{equation2}
\end{equation}
By controlling  the value of $\langle v^2 \rangle$ we can reach a
chosen temperature $T_f$. An appropriated way to do that is by
successive approximations. We chose a set of velocities $\{v \}_0$
so that we get $\langle v^2 \rangle_0$. By multiplying  each
velocity by a factor $\alpha_0$ defined as
\begin{equation}
\alpha_0 = \sqrt{\frac{m}{3k_B}\frac{{\langle v^2 \rangle}_0}{T_f}} ~~~,
\label{equation3}
\end{equation}
a first approximation to $T_f$ is done. By evolving in time the
system we can create sequences, $T_n$, $\{v \}_n$ and $\{\alpha
\}_n$, such that after a finite number of time steps the
temperature of the system converges to $T_n \approx T_f$. The
friction coefficient is calculated as the quotient
\begin{equation}
\mu \equiv \frac{dF_x}{dF_z} .
\end{equation}

\begin{figure}\vspace{0.4cm}
  \includegraphics[width=4.0cm]{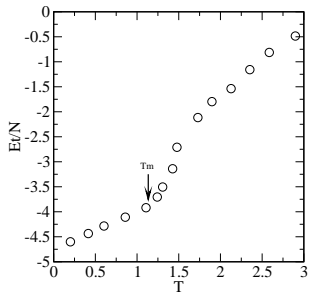}\\
  \caption{\scriptsize{Energy as a function of temperature.
  The melting temperature is estimated as the inflexion
  point, being around $T_m\approx1.1 \epsilon/k_b$}}
  \label{EtxT}
\end{figure}

Before we start the simulation we have to have an estimative of the
melting temperature, $T_m$, of the system. This is done by
performing a preliminary simulation of the substrate. In Figure
\ref{EtxT} we show the total energy per particle, $E$, as a function
of temperature. The melting temperature is estimated as the
inflection point of the curve. We find $T_m \approx 1.1$ in
accordance with earlier calculations
\cite{pablo1,pablo2,flavio1,flavio2}.
%
%
The velocity, position and forces are stored at each time step for
further analysis. We measure the time $t$, temperature $T$ and
forces in units of $\sigma  \sqrt{m/\epsilon}$, $\epsilon/k_B$ and
$\epsilon/\sigma$ respectively.
%
%
\section{Results \label{section3}}
\begin{figure}
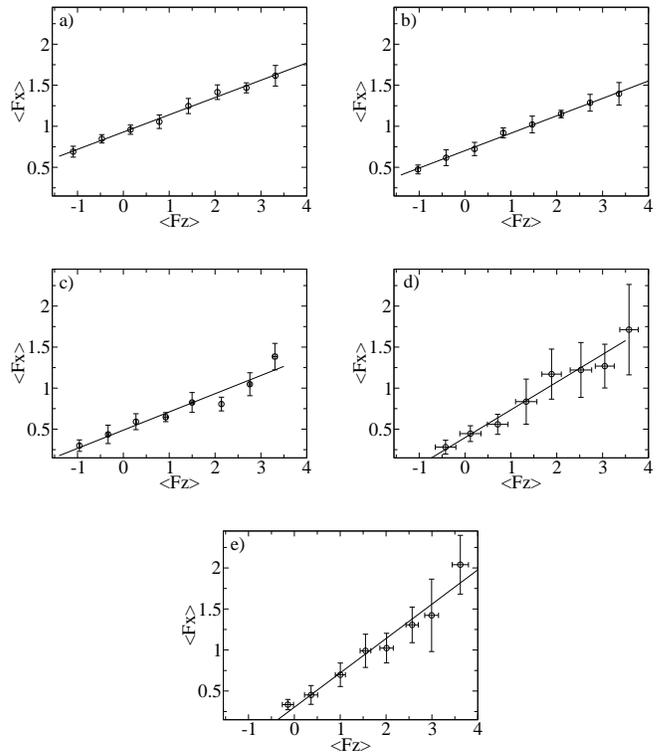
\vspace{0.25cm}
\includegraphics[width=4.0cm]{FxxFzT1.eps}~~~~
\includegraphics[width=4.0cm]{FxxFzT2.eps}\\
\vspace{0.5cm}
\includegraphics[width=4.0cm]{FxxFzT3.eps}~~~~
\includegraphics[width=4.0cm]{FxxFzT4.eps}\\
\vspace{0.5cm}
\includegraphics[width=4.0cm]{FxxFzT5.eps}\\
\caption{\footnotesize {The friction force,$\langle F_x \rangle$
as a function of normal force, $\langle F_z \rangle$ is shown for several
temperatures. The $\langle F_x \rangle$ and $\langle F_z \rangle$
forces are measured in units of $\epsilon/\sigma$. The figures,
from $a$ to $e$ are for several different values of $T=0.25, 0.44, 067, 085, 1.05$
respectively. The circles are the MD results and the straight line
correspond to an adjust.}}
\label{FlxFzT1to5}
\end{figure}

\begin{figure}
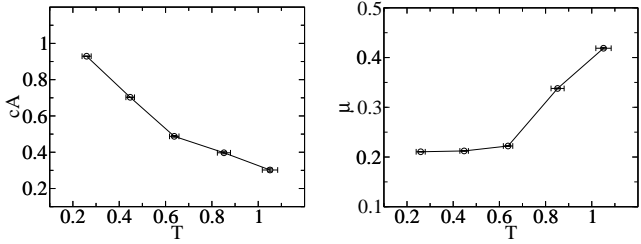
\vspace{0.25cm}
\includegraphics[width=4.0cm]{cAxT.eps}~~~~\includegraphics[width=4.0cm]{muxT.eps}\\
\caption{\footnotesize {Plot of $cA$ (left) and $\mu$
(right) as a function of Temperature. The line is only a
guide to the eyes.}} \label{AreaxT}
\end{figure}

\begin{figure}
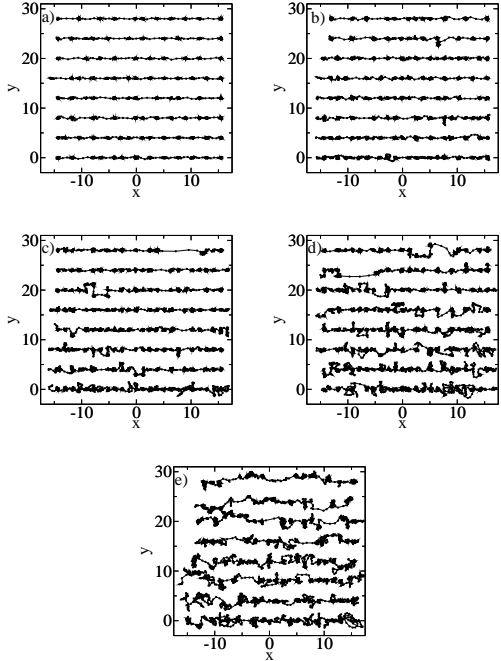
\vspace{0.25cm}
\includegraphics[width=3.0cm]{PlanoXYT1.eps}~~~~
\includegraphics[width=3.0cm]{PlanoXYT2.eps}\\
\vspace{0.45cm}
\includegraphics[width=3.0cm]{PlanoXYT3.eps}~~~~
\includegraphics[width=3.0cm]{PlanoXYT4.eps}\\
\vspace{0.45cm}
\includegraphics[width=3.0cm]{PlanoXYT5.eps}\\
\caption{\footnotesize {Path of the tip over the surface (XY
plane) for several temperatures and normal forces. From a) to e)
we have $T=0.25, 0.44, 067, 085, 1.05$ respectively.
The normal forces are defined in fig \ref{FlxFzT1to5}.
The plots are shown dislocated by a constant value in the $y$ direction
as a matter of clarity.}} \label{planoXY}
\end{figure}

We have simulated the $FFM$ system for several temperatures and
initial distances of the tip to the substrate or equivalently, the
normal force in the tip. In figure \ref{FlxFzT1to5} we show a plot
of our MD simulation results for the friction force as a function of
normal force for several temperatures.
%
%
%
%
The Amonton's Law of friction states that frictional forces are
proportional to the normal force and independent of the contact
area. This type of behavior was observed in some systems by many
authors, who fitted $\langle F_x \rangle$ to a linear function of
both load,$\langle F_z \rangle$, and contact area, $A$:
\begin{equation}\label{linearfit}
\langle F_x \rangle =\mu \langle F_z \rangle + cA.
\end{equation}
Here $\mu$ is the friction coefficient and the second term $cA$ is
interpreted as the friction force for zero normal force. Ringlein
et. al.\cite{Ringlein} showed that the friction forces in
nanometric adhesive systems depends on the contact area, breaking
down the Amonton's law. Because we use the LJ force between the
tip and the surface is adhesive. In the following we present our
results that strongly suggest that the Amonton's
laws\cite{Ringlein,Gourdon} is violated when the friction force is
considered as a function of temperature. In the figure
\ref{AreaxT}(left) we show a plot of $cA$ as a function of
temperature. We can see that when the temperature increases the
contact area or adhesion forces decrease. This behavior can be
related to the fact that at low temperature the atoms at the
surface perform low amplitude jumps so that the number of
collisions with the tip is low. In this case the effective contact
area is high because the tip stays a long time close to the
surface. However, when the temperature grows the number of high
energy fluctuations of particles at the surface increases with a
consequent increase in the number of collisions of high energy
with the tip, decreasing the effective contact area.
%
%
We also observe that the friction coefficient (Shown in figure
\ref{AreaxT}(right)) grows abruptly  at $T\sim 0.7
\varepsilon/k_b$. In the following we discuss the fact that this
behavior can be related to the pre-melting of the surface. We show
in fig. \ref{planoXY} a plot of the path of the tip over the
surface for several temperatures and normal forces.
%
%
%
%
As should be expected the paths are well defined for low
temperatures becoming random as temperature grows.
%
\begin{figure}
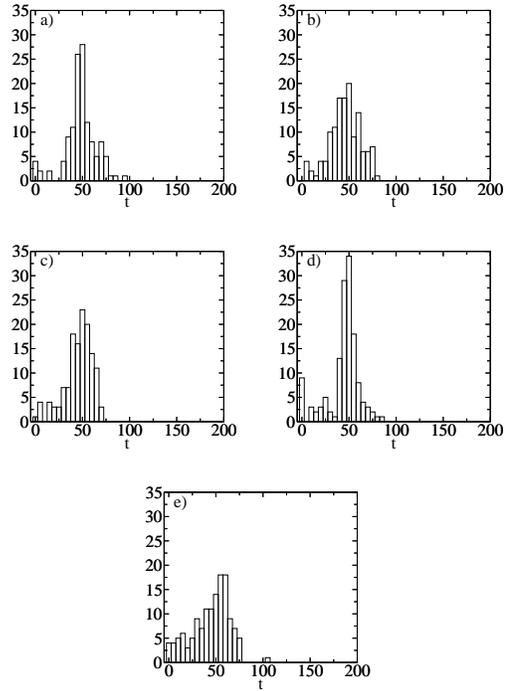
\vspace{0.25cm}
\includegraphics[width=3.0cm]{HistoNor1-T1.eps}~~~~
\includegraphics[width=3.0cm]{HistoNor1-T2.eps}\\
\vspace{0.45cm}
\includegraphics[width=3.0cm]{HistoNor1-T3.eps}~~~~
\includegraphics[width=3.0cm]{HistoNor1-T4.eps}\\
\vspace{0.45cm}
\includegraphics[width=3.0cm]{HistoNor1-T5.eps}\\
\caption{\footnotesize {Histogram for the residence time. The
normal force is for $F_z=-1.09,-1.02, -0.95, -0.42, -0.14$. Beam
size is $t_{beam}=5$}.} \label{HistT02-10N1}
\end{figure}
%
%
\begin{figure}
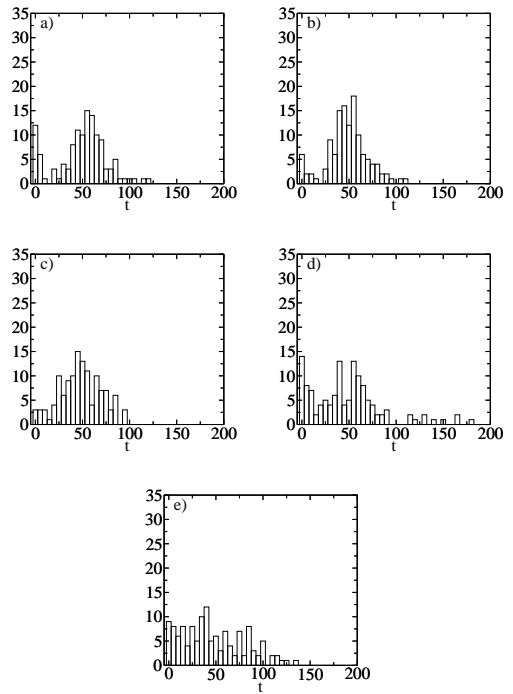
\vspace{0.25cm}
\includegraphics[width=3.0cm]{HistoNor6-T1.eps}~~~~
\includegraphics[width=3.0cm]{HistoNor6-T2.eps}\\
\vspace{0.45cm}
\includegraphics[width=3.0cm]{HistoNor6-T3.eps}~~~~
\includegraphics[width=3.0cm]{HistoNor6-T4.eps}\\
\vspace{0.45cm}
\includegraphics[width=3.0cm]{HistoNor6-T5.eps}\\
\caption{\footnotesize {Histogram for the residence time. The
normal force is for $F_z=2.05, 2.09, 2.13, 2.52, 2.56$. Beam size
is $t_{beam}=5$}.} \label{HistT02-10N6}
\end{figure}
To understand what is going on, we have calculated the residence
time of the tip in each site, defined as the time that the tip spent
in the neighborhood of a specific site, i.e., its distance to some
particular site being smaller than some reference distance $\delta$.
With no loss of generality we choose $\delta=\sigma$, the lattice
parameter. In figure \ref{HistT02-10N1} and \ref{HistT02-10N6} we
show the histograms for normal forces $F_z=-1.09,-1.02, -0.95,
-0.42, -0.14$ and $F_z=2.05, 2.09, 2.13, 2.52, 2.56$ respectively
for several temperatures. For negative values of $F_z$ the residence
time is well defined even at high temperatures having its average at
$\delta t \approx 50$. The tip is immersed in the surface, so that,
it can easily travel along channels on the surface of the crystal.
In the figure \ref{HistT02-10N6} we show the histograms for positive
values of $F_z$. The temperatures are the same as in figure
\ref{FlxFzT1to5}. At low T they show a similar behavior as that for
$F_z < 0 $. However, at higher temperatures, the residence time
spreads out to the higher $t$ region. We interpret this as a closing
of the channels discussed above due to thermic motion of the
particles at the surface. When temperature increases the particle
gets more energy, which is eventually enough to push it from any
specific neighborhood.
\section{Conclusion \label{section4}}
We have performed a molecular dynamics simulation of a $FFM$
experiment. Our results were obtained by varying the normal force in
the tip and the temperature of the surface. The behavior of the $cA$
term in the Amonton's law (eq.

\ref{linearfit}) and the friction coefficient were found to depend
on the temperature. The $cA$ term which measures essentially the
effective contact area, $A$, between the tip and the surface were
found to decrease with increasing $T$. The friction coefficient
presents a sudden jump which seems to be related to the premelting
processes of the surface.
\acknowledgements This work was partially supported by CNPq. We
are grateful to B.A. Soares for suggestions and comments.
%
%

%
\end{document}